\documentclass{article}
\usepackage[preprint]{spconfa4}
\usepackage{amsmath,graphicx}

\usepackage{tikz}
\usetikzlibrary{arrows.meta,shapes.arrows,chains,decorations.pathreplacing}

\usetikzlibrary{plotmarks}

\usepackage{amsmath}
\usepackage{algorithm}

\usepackage{enumitem}

\graphicspath{ {Images/} }
\usepackage{amsmath,amssymb,amsfonts}
\usepackage{multirow}
\usepackage{algorithmic}
\usepackage{graphicx}
\usepackage{textcomp}
\usepackage{xcolor}
\def\BibTeX{{\rm B\kern-.05em{\sc i\kern-.025em b}\kern-.08em
    T\kern-.1667em\lower.7ex\hbox{E}\kern-.125emX}}
    
\usepackage{url}

\usepackage[noadjust]{cite}

\copyrightnotice{978-1-6654-6867-1/22/\$31.00~\copyright2022 IEEE}
                   
\title{GMM based Multi-stage Wiener Filtering for Low SNR Speech Enhancement}
\name{Wageesha~Manamperi, 
Prasanga~N.~Samarasinghe, 
Thushara~D.~Abhayapala and
Jihui~Zhang}
\address{Audio \& Acoustic Signal Processing Group, The Australian National University, Canberra, Australia}
\begin{document}
%
\maketitle
\begin{abstract}
This paper proposes a single-channel speech enhancement method  to reduce the noise and enhance speech at low signal-to-noise ratio (SNR) levels and non-stationary noise conditions. Specifically, we focus on modeling the noise using a Gaussian mixture model (GMM) based on a multi-stage process with a parametric Wiener filter. The proposed noise model estimates a more accurate noise power spectral density (PSD), and allows for better generalization under different noise conditions such as harmonic and babble noise environments compared to traditional Wiener filtering methods. Simulations show that the proposed approach can achieve better performance in terms of speech quality (PESQ) and intelligibility (STOI) at low SNR levels.
\end{abstract}
%
\vspace*{-0.4cm}

\section{Introduction}
\label{sec:1} 
\vspace*{-0.2cm}
Speech enhancement is useful in many applications such as automatic speaker recognition, voice communication, and hearing aid devices  \cite{wang2018supervised}. Notable contributions to speech enhancement were developed over the past few decades. However, the speech enhancement in low signal-to-noise ratio (SNR) conditions is still challenging. This paper addresses the problem of single-channel (i.e., one-microphone) speech enhancement using a Wiener filtering approach for low SNR levels and non-stationary noise conditions.
    
Single-channel speech enhancement methods can be classified mainly into statistical-based approaches and data-driven approaches. Statistical-based approaches include spectral subtraction \cite{loizou2007speech}, Wiener filtering \cite{scalart1996speech,srinivasan2005codebook, chehrehsa2016single,chehresa2011mmse,savoji2014speech}, the  minimum mean-square error (MMSE) short-time spectral amplitude estimation \cite{ephraim1984speech}, etc. In general, these methods tend to perform poorly in non-stationary noisy conditions and thus often assume stationarity to obtain good performance. Compared to data driven methods, statistical methods are computationally efficient, yet the overall quality of their speech enhancement performance is typically lesser.

Existing data-driven approaches for speech enhancement include non-negative matrix factorization (NMF) \cite{mohammadiha2013supervised} and neural networks (NNs) \cite{wang2014training, wang2005ideal, williamson2015complex, choi2018phase, pascual2017segan, rethage2018wavenet, luo2019conv, pandey2019tcnn, zeghidour2021wavesplit, saleem2019deep}. They often use learning methods with multi-layers representations with large training sets of data to achieve high quality enhancement. Prior work on deep-learning based methods have recently attracted much attention. Supervised learning methods using deep neural networks (DNNs) can be classified into two categories mainly as Time-Frequency (TF) domain methods and time domain methods. Generally, the DNN-based TF algorithms estimate a real or complex value mask for each TF bin such as ideal binary mask (IBM) and ideal ratio mask (IRM) to obtain the clean speech spectra and show robustness to non-stationary noise sources \cite{wang2014training, wang2005ideal, williamson2015complex, choi2018phase}. 
The time domain DNN-based algorithms such as SEGAN \cite{pascual2017segan}, WaveNeT \cite{rethage2018wavenet}, Conv-TasNet \cite{luo2019conv}, TCNN \cite{pandey2019tcnn} and Wavesplit \cite{zeghidour2021wavesplit} have shown great improvements over TF methods. These techniques model its clean speech equivalent accurately in high ($>0$ dB) and moderate ($>-5$ dB) SNR levels using large models with high complexity and long hours of training and validation data.


In our contribution, an extension of the Gaussian mixture model (GMM) based single-channel speech enhancement approach in \cite{chehresa2011mmse} for modeling noise power spectral densities (PSDs) is presented. We improve the performance through a multi-stage parametric Wiener filter at low SNR levels and non-stationary noise conditions. Motivated by the success of the Wiener filtering methods in noise reduction, we propose to model noise in a multi-stage process using a set of Gaussian components and a parametric Wiener filter. In this way, the spectral enhancement is achieved by considering the set of noise GMM mean vectors based on its energy level at each stage. The proposed method is a simple, efficient, and unsupervised approach that improves speech quality while preserving speech intelligibility in noisy environments. We will show that multi-stage speech enhancement can improve performance by providing more accurate estimates of the noise PSD. Simulations show that a better speech enhancement is achieved compared to the Wiener filtering method in \cite{chehresa2011mmse}. 

\vspace*{-0.4cm}
\section{\label{sec:2} Problem Definition}

\vspace*{-0.2cm}
We address the single-channel speech enhancement problem by focusing on the case of only two sources, speech and noise. Let the received noisy speech be 
\[
      m(f,t) = s(f,t) + v(f,t),
\]
where $s(f,t)$ and $v(f,t)$ are the Short Time Fourier Transforms (STFTs) of the desired speech and noise, respectively, $t$ and $f$ denotes the time frame and frequency bin indices, respectively with $t \in \{t_1,\dotsc,t_T\}$ and $f \in \{f_1,\dotsc,f_F\}$ where $T$ and $F$ are the number of time frames and frequency bins, respectively.

Our problem is to extract the speech signal $s(f,t)$ from the noisy speech signal $m(f,t)$ under low SNR conditions.
 
\vspace*{-0.4cm}

\section{Parametric Wiener filtering  \label{sec:3}}
\vspace*{-0.2cm}
Due to its simplicity, the Wiener filter has been used for noise reduction in the past. As one of the steps in our novel GMM based noise suppression framework, we propose a parametric Wiener filter in this section. 

We can use the typical Wiener filtering to estimate the speech in the STFT domain as $\hat{s}(f,t) = w(f,t)m(f,t)$,
where $w(f,t)$ is the gain function of the Wiener filter 
\begin{equation}\label{eq:wiener}
   w(f,t) = \frac{ {\Phi}_s(f,t)}{{\Phi}_s(f,t) + {\Phi}_v(f,t) },
\end{equation}
%
where ${\Phi}_s(f,t) \triangleq \text{E}\{|s(f,t)|^2\}$ and ${\Phi}_v(f,t) \triangleq \text{E}\{|{v}(f,t)|^2\}$ denote the power spectral density (PSD)  of the speech and noise, respectively, at the $f^{\text{th}}$ frequency bin of frame $t$. 

Since a parametric Wiener filter allows a compromise between the amount of speech distortion and noise reduction and is known to perform better in the presence of non-stationary noise conditions \cite{haykin2008adaptive}, we introduce a parameter to the noise variance $\beta$ and a power coefficient $\gamma$ that imposes a constant on the maximum allowable distortion of the desired speech signal and provides the minimum mean squared error (MMSE) linear estimator of ${s}(f,t)$ \cite{lim1979enhancement} as
\begin{equation}\label{eq:PWF}
     \hat{s}(f,t) = \Big(\frac{ {\Phi}_s(f,t)}{{\Phi}_s(f,t) + \beta {\Phi}_v(f,t) }\Big) ^ \gamma m(f,t).
\end{equation}
Here, $\beta$ is the over- or underestimation factor and controls the compromise between noise reduction and speech distortion. 
$\beta$ can be fixed or frequency dependent on the spectral properties of the desired speech. However, in this work, we mainly focus on fixed values of $\beta$ for more aggressive noise reduction.

\vspace*{-0.4cm}
\section{GMM based Speech Enhancement \label{sec:4}}
\vspace*{-0.2cm}
\subsection{Modeling speech and noise PSDs with GMMs \label{subsec:4.1}}
\vspace*{-0.2cm}
We use the Gaussian Mixture Model method \cite{vaseghi2008advanced} to model the PSDs of the mixture, speech and noise using the expectation-maximization (EM) algorithm. 
A GMM for a mixture signal can be composed of $K$ Gaussians, and is given by
\[
\Gamma_m(f) = \sum_{k=1}^K C_{mk} G_{mk}(f;\mu_{mk},\Sigma_{mk}),
\]
where $\Gamma_m(\cdot)$ is the probability density function (PDF) of the mixture signal when both speech and noise signals are present, $C_{mk}$ is the contribution of the $k^{th}$ GMM component in the mixture recording, and $G_{mk}$ is the Gaussian mixture of the mixture recording for the $k^{th}$ GMM component and $f^{\text{th}}$ frequency. Also, $\mu_{mk}$ and $\Sigma_{mk}$ denote the mean and covariance matrix of the mixture for the $k^{th}$ GMM component and $f^{th}$ frequency, respectively.
Likewise, we can write the GMM of clean speech as 
\[
\Gamma_s(f) = \sum_{k=1}^{K_s} C_{sk} G_{sk}(f;\mu_{sk},\Sigma_{sk}),
\]
and the GMM of noise as 
\[
\Gamma_v(f) = \sum_{k=1}^{K_v} C_{vk} G_{vk}(f;\mu_{vk},\Sigma_{vk}),
\]
where $K_s$ and $K_v$ denotes the number of GMM components for the speech and noise, respectively. The means and covariances are denoted by $\mu$, and $\Sigma$ with the speech and noise as a suffix and $\mu_{sk}$ and $\mu_{vk}$ are trained offline in the training phase.

We define the GMM mean vector of speech as $\boldsymbol{\mu}_{sk} = [ \mu_{sk}(f = f_1), \mu_{sk}(f = f_2), \ldots, \mu_{sk}(f = f_{F}) ]^T $ by arranging the mean values of each frequency for a single $k^{th}$ GMM component into a vector. Similarly, we can define the GMM mean vector for noise as $\boldsymbol{\mu}_{vk} =  [ \mu_{vk}(f = f_1), \mu_{vk}(f = f_2), \ldots, \mu_{vk}(f = f_{F}) ]^T$.

\vspace*{-0.4cm}
\subsection{Multi-stage parametric Wiener filtering \label{subsec:4.2}}
\vspace*{-0.2cm}
The proposed model (as summarized in Algorithm \ref{alg:PWF}), has a {\em training phase} and an {\em enhancement phase}. During the training phase, we extract the GMM mean vectors of clean speech and noise PSDs which will be fed into the enhancement phase.

We propose a multi-stage parametric Wiener filter method to improve the speech quality in the enhancement phase. At each stage $i = 1, \ldots, I$,  we update a set of noise GMM mean vectors $\boldsymbol{\mu}^{(i)}_{vk}$ to obtain a better estimate of noise PSD and  Wiener filtering gain. We determine the number of stages $I$ by choosing a set of $\mu^{(i)}_{vk}(f)$ based on their energy level at each stage $i$. Basically, we consider the high energy noise GMM mean vectors in the first stage, and group the rest in the subsequent stages.

We write the estimated PSD  of noise as
%
\begin{equation} \label{eqn:PSDnoise}
{\Phi}^{(i)}_v(f,t) = \sum_{k=1}^{K^i_{v}} \mu^{(i)}_{vk}(f)  \, \alpha^{(i)}_{vk}(t),
\end{equation}
%
where $K^i_{v}$ denotes the number of noise GMMs in stage $i$, $\mu^{(i)}_{vk}(f)$ and $\alpha^{(i)}_{vk}(t)$ are GMM mean values and power coefficients for the noise at stage $i$, respectively. 

The estimated PSD of speech can be written as 
\begin{equation} \label{eqn:PSDspeech}
{\Phi}_s(f,t) = \sum_{k=1}^{K_s} \mu_{sk}(f) \,  \alpha_{sk}(t),
\end{equation}
where $\alpha_{sk}(t)$ is the power coefficients corresponding to the GMM mean values 
for the speech. 

Assuming the speech signal and noise are uncorrelated, we can write the PSD of the mixture as the weighted sum of the GMM means of both speech and noise
%
\begin{equation} \label{eqn:PSD_estimate}
\hat{{\Phi}}_m(f,t) = \sum_{k=1}^{K_s} \mu_{sk}(f)  \alpha_{sk}(t) + \sum_{k=1}^{K^{(i)}_{v}} \mu^{(i)}_{vk}(f)  \alpha^{(i)}_{vk}(t).
\end{equation}
Writing \eqref{eqn:PSD_estimate} for each  frequency bin $f$, we have a set of linear equations in a matrix form for the time frame $t$ as
\begin{equation} \label{eqn:PSD_Matrix}
\boldsymbol{\hat{\boldsymbol{\Phi}}}_m(t) = 
\begin{bmatrix}
\boldsymbol{U}_{s}
\,
\boldsymbol{U}^{(i)}_{v}
\end{bmatrix}
\begin{bmatrix}
\boldsymbol{\alpha}_{s}(t)\\
\boldsymbol{\alpha}^{(i)}_{v}(t)
\end{bmatrix}
\end{equation}
where $\boldsymbol{U}_{s} =[\boldsymbol{\mu}_{s1}, \dots, \boldsymbol{\mu}_{sK_s} ]$, $\boldsymbol{U}^{(i)}_{v} =[\boldsymbol{\mu}^{(i)}_{v1}, \ldots, \boldsymbol{\mu}^{(i)}_{vK_v^{(i)}}]$, $\boldsymbol{\alpha}_{s}(t)=[\boldsymbol{\alpha}_{s1}(t), \ldots, \boldsymbol{\alpha}_{sK_s}(t)]^T$ and  $\boldsymbol{\alpha}^{(i)}_{v}(t)=[\boldsymbol{\alpha}^{(i)}_{v1}(t), \ldots, \\ \boldsymbol{\alpha}^{(i)}_{vK_v^{(i)}}(t)]^T$. 
We can solve \eqref{eqn:PSD_Matrix} to find the coefficients of the current time frame based on the mixture PSD at stage $i$
%
\begin{equation} \label{eqn:alphacoeff}
   \begin{bmatrix}
\boldsymbol{\alpha}_{s}(t)\\
\boldsymbol{\alpha}^{(i)}_{v}(t)
\end{bmatrix}  = \begin{bmatrix}
\boldsymbol{U}_{s}
\,
\boldsymbol{U}^{(i)}_{v}
\end{bmatrix}^{\ddagger}  \boldsymbol{\hat{\boldsymbol{\Phi}}}_m(t)
\end{equation}
%
where $\cdot^\ddagger$ is the Moore-Penrose inverse.

This estimation may cause some power coefficients to be negative due to the pseudo-inverse of the mean matrix $\begin{bmatrix}
\boldsymbol{U}_{s}
\,
\boldsymbol{U}^{(i)}_{v}
\end{bmatrix}$. The negative power coefficients make PSDs negative which is not possible. To avoid that, we normalize the matrix to set all the values to be positive such that the negative power coefficients are set to zero and the total power of the matrix is the same as the original matrix. We can reconstruct an estimation of the speech-only PSD and noise-only PSD in the current time frame using the (\ref{eqn:PSDspeech}) and (\ref{eqn:PSDnoise}), respectively.

Using \eqref{eq:PWF}, we construct the transfer function of the parametric Wiener filter  $w^i(f,t)$, and use it to filter the STFT of the mixture 
and obtain the enhanced speech at stage $i$ as
\[
 \hat{s}^i(f,t) = w^i(f,t)\,m(f,t).
\]
Using the inverse STFT, $\hat{s}^i(f,t)$ can be converted to $\hat{s}^i(t)$ as the enhanced speech at the $i^{th}$ stage. We continue to filter out  $\hat{s}^i(t)$ at the next stage $i+1$  by considering a set of noise GMM mean vectors $\mu^{i+1}_{vk}$.

\begin{algorithm}[htb]
\caption{Multi-stage speech enhancement}\label{alg:PWF}
\begin{enumerate} [noitemsep]
    \item{\textbf{Training phase}
    \begin{itemize} [noitemsep]
        \item{STFT of the clean speech, and noise-only signal: $s(f,t)$, and $v(f,t)$.}
        \item{Model ${\Phi}_s(f,t)$, and ${\Phi}^{(i)}_v(f,t)$ with GMMs.}
        \item{Extract ${\mu}_{sk}$ and ${\mu}_{vk}$.}
    \end{itemize}
    }
    \item{\textbf{Enhancement phase}
    \begin{itemize} [noitemsep]
        \item{STFT of the noisy signal $m(f,t)$.}
        \item{Select a set of $\mu^{(i)}_{vk}(f)$ based on $I$.}
        \item{Update filters for stage $i \in I $: 
        \begin{itemize}
            \item{use $\mu_{sk}(f)$ and $\mu^{(i)}_{vk}(f)$, and estimate $\hat{\boldsymbol{\Phi}}_s(f,t)$ $\hat{\boldsymbol{\Phi}}^{(i)}_v(f,t)$}
            \item{Estimate $\hat{s}^i(f,t)$ using the parametric Wiener filtering with $\beta$, and $\gamma$}
        \end{itemize}
        }
        \item{iSTFT of the enhanced signal $\hat{s}^I(f,t)$.}
    \end{itemize}
    }
\end{enumerate}
\end{algorithm}

\setlength{\textfloatsep}{5pt}

\vspace*{-0.4cm}
\section{\label{sec:5} Experiment Validation}
\vspace*{-0.3cm}
\subsection{\label{subsec:5:1}  Implementation Details}
\vspace*{-0.2cm}
\subsubsection{Experimental Setup}
\vspace*{-0.2cm}
We generate the noisy speech by adding clean speech to noise with SNR levels of $-10$ dB, $-5$ dB, $0$ dB, $5$ dB, and $10$ dB. In the training phase, we use a clean speech dataset from the TIMIT database \cite{garofolo1993timit} and a noise dataset from the NOISEX database \cite{Noisex} to model the GMMs of speech and noise PSDs, respectively. Also, here we normalize the PSDs to make their energy equal to one.
We use two different types of noise: multi-talker {\em babble noise} and {\em destroyer engine noise} from the NOISEX database. It should be noted that we use different speech data to construct the noisy signal for the enhancement phase. 

\vspace*{-0.4cm}
\subsubsection{Algorithm Parameters}
\vspace*{-0.2cm}
The number of Gaussians for the speech and noise GMMs are chosen to be $K_s = 6$ and $K_v = 9$ following a trial and error approach proposed in \cite{chehrehsa2016single},\cite{chehresa2011mmse}. In order to obtain better denoising performance at low SNR levels, we use $\beta = 2 $ and $\gamma = 1$. We set $I=2$ and all signals are down sampled at 8 kHz and transformed to the frequency domain by a 512 point STFT using a 20 ms Hanning window with zero padding and 50\% of overlap.

\vspace*{-0.4cm}
\subsubsection{Evaluation Methods}
\vspace*{-0.2cm}
We evaluate the quality of the enhanced speech using objective measures, (i) perceptual evaluation of speech quality (PESQ) \cite{recommendation2001perceptual} and (ii) the short-time objective intelligibility score (STOI) \cite{taal2011algorithm}. The range of PESQ score is from $-0.5$ to $4.5$, whereas STOI score is from $0\%$ to $100\%$. 

\vspace*{-0.4cm}
\subsection{\label{subsec:5:2} Results and Analysis}
\vspace*{-0.2cm}
\begin{figure}[h!]
\centering
\begin{tikzpicture}[font=\scriptsize]
\node[above right] (img) at (0,0)
{
\includegraphics[width=3.3in]{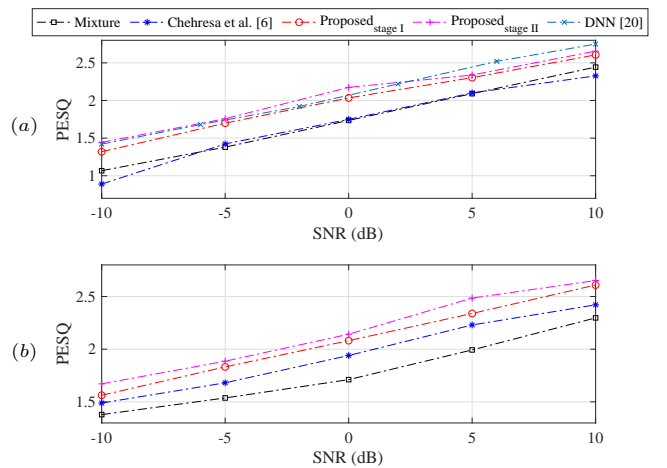}
};
\node[text=black] at (0.2,4.8) {$(a)$};
\node[text=black] at (0.2,1.8) {$(b)$};
\end{tikzpicture}
\vspace*{-0.8cm}
\caption{ \label{fig:pesq} PESQ scores for (a) babble noise, and (b) destroyer engine noise.}
\end{figure}  

\vspace{0.3cm}

\begin{figure}[htb]
\centering
\begin{tikzpicture}[font=\scriptsize]
\node[above right] (img) at (0,0)
{
\includegraphics[width=3.3in]{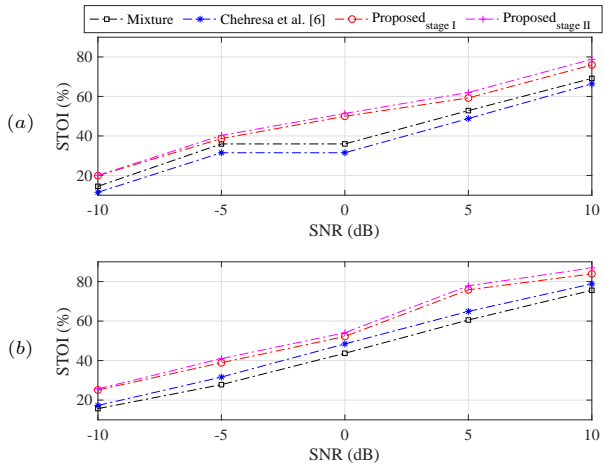}
};
\node[text=black] at (0.2,4.8) {$(a)$};
\node[text=black] at (0.2,1.8) {$(b)$};
\end{tikzpicture}
\vspace*{-0.8cm}
\caption{ \label{fig:stoi} STOI scores (in percent) for (a) babble noise, and (b) destroyer engine noise.}
\end{figure}  

\setlength{\textfloatsep}{5pt}

We compare our approach with Chehresa et al. \cite{chehresa2011mmse}. Figure \ref{fig:pesq} and \ref{fig:stoi} show the speech enhancement performance over the two noise types. We observe that the proposed method can achieve higher PESQ and STOI scores at all input SNRs compared to  \cite{chehresa2011mmse} for both babble and engine noise. In the proposed algorithm, first stage does a considerable improvement while second stage does a minor improvement. In total it is a significant PESQ improvement compared to \cite{chehresa2011mmse}. The proposed method leads to a significant improvement of predicted intelligibility, and a near-constant PESQ improvement of about 0.4 over the \cite{chehresa2011mmse} and about 0.3 over the mixture signal for babble noise. Whereas PESQ improves about 0.2 over the \cite{chehresa2011mmse} and about 0.4 over the mixture signal for destroyer engine noise. This suggests the proposed method improves speech quality without degrading intelligibility. Informal listening tests confirm these results. Furthermore, we evaluate the proposed method with DNN method (results from \cite{saleem2019deep} which uses the same dataset \cite{Noisex}) in terms of PESQ score for babble noise in Fig.~\ref{fig:pesq} (a). The PESQ scores for babble noise are similar to the proposed method and DNN at low SNR levels until around $3$ dB. However, the DNN method has slightly improved PESQ scores at high SNR conditions. Note that we do not show the result of the STOI score of the DNN method due to its unavailability in the reference paper. 

\begin{figure}[ht]
\centering
\begin{tikzpicture}[font=\scriptsize]
\node[above right] (img) at (0,0)
{
\includegraphics[scale=0.47]{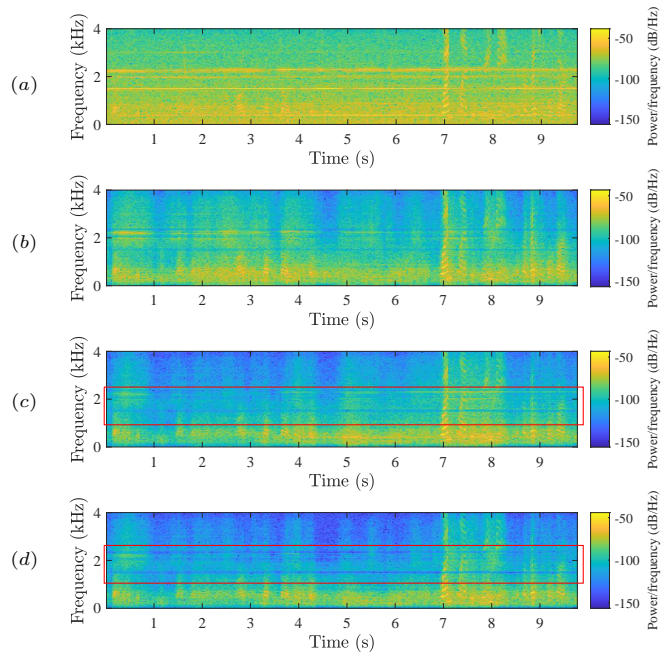}};
\node[text=black] at (0.2,8.4) {$(a)$};
\node[text=black] at (0.2,6.3) {$(b)$};
\node[text=black] at (0.2,4.2) {$(c)$};
\node[text=black] at (0.2,2.1) {$(d)$};
\draw[red] (1.25,2.3) -- (7.55,2.3) -- (7.55,1.8) -- (1.25,1.8) -- (1.25,2.3);
\draw[red] (1.25,4.4) -- (7.55,4.4) -- (7.55,3.9) -- (1.25,3.9) -- (1.25,4.4);
\end{tikzpicture}
\vspace*{-1.5cm}
\caption{ \label{fig:spectrograms}Spectrogram plots of (a) the mixture, (b) Chehresa et al. \cite{chehresa2011mmse}, (c) stage I, and (d) stage II of the proposed method for {\em destroyer engine noise} at SNR level of $-5$ dB.}
\end{figure}

To further understand the staging effect in the proposed method, we examine the time varying spectrograms for the destroyer engine noise source at SNR level of $-5$ dB in Fig.~\ref{fig:spectrograms}. Mixture signal is the clean speech contaminated by engine noise, and has PESQ of $1.54$ and STOI of $27.82\%$ scores. We notice that the destroyer engine noise has a higher harmonic structure and those GMM mean vectors lie in between $0 - 2375$ Hz. For \cite{chehresa2011mmse}, we observe that the large residual noise is evident in Fig.~\ref{fig:spectrograms}(b). Whereas Fig.~\ref{fig:spectrograms}(c) and (d) show the noise reduction of the proposed dual staging process in the frequency region over $0$ to $4$kHz. In order to highlight the noise suppression at each stage, we show a rectangle box on the plots (c) and (d). Fig.~\ref{fig:spectrograms}(d) shows the enhanced signal has PESQ of $1.89$ and STOI of $41\%$ scores. Note that we use only two stages ($I=2$) in the simulations to obtain the enhanced signal as the first stage performs a significant improvement compared to the second stage. However, the number of stages in the proposed method for an optimal output under a given noisy environment is left for future work.  

\vspace{-0.5cm}

\vspace{-0.3cm}
Overall, the proposed method achieves noise suppression without compromising speech quality. Also, the dual staging method allows us to obtain a more precise estimate of the Wiener filter gain compared to the traditional Wiener filter. It also seems that \cite{saleem2019deep} DNN method performance with babble noise is slightly similar to the proposed method at low SNR conditions.


\vspace{-0.3cm}
\section{\label{sec:6} Conclusion}
\vspace{-0.2cm}
In this paper, we have presented a novel single-channel speech enhancement approach using a GMM based multi-stage parametric Wiener filtering method at low SNR conditions. We show the proposed method facilitates dynamic noise adaptation and improves the speech quality while preserving speech intelligibility in a non-stationary noise environment with a simple multi-stage parametric Wiener filter. Evaluations verify that the proposed method achieves a better speech enhancement and noise suppression compared to the conventional Wiener filtering methods in terms of PESQ and STOI. For future work, we will investigate the applicability of the proposed method and its influence on objective speech quality at very low SNR levels.

\twocolumn
\vspace{-0.2cm}
\bibliographystyle{IEEEbib}
{\fontsize{9}{11}
\bibliography{reference}}

\begin{thebibliography}{10}

\bibitem{wang2018supervised}
D.~Wang and J.~Chen,
\newblock ``Supervised speech separation based on deep learning: An overview,''
\newblock {\em IEEE Trans. on Audio, speech, and Lang. Process.}, vol. 26, no.
  10, pp. 1702--1726, May 2018.

\bibitem{loizou2007speech}
P.~C. Loizou,
\newblock {\em Speech enhancement: theory and practice},
\newblock CRC press, 2007.

\bibitem{scalart1996speech}
P.~Scalart,
\newblock ``Speech enhancement based on a priori signal to noise estimation,''
\newblock in {\em Proc. IEEE Int. Conf. on Acoust., Speech and Signal
  Process.}, May 1996, vol.~2, pp. 629--632.

\bibitem{srinivasan2005codebook}
S.~Srinivasan, J.~Samuelsson, and W.~B. Kleijn,
\newblock ``Codebook driven short-term predictor parameter estimation for
  speech enhancement,''
\newblock {\em IEEE Trans. on Audio, speech, and Lang. Process.}, vol. 14, no.
  1, pp. 163--176, Dec. 2005.

\bibitem{chehrehsa2016single}
S.~Chehrehsa,
\newblock {\em Single-Channel Speech enhancement using statistical modelling},
\newblock Ph.D. thesis, Auckland University of Technology, 2016.

\bibitem{chehresa2011mmse}
S.~Chehresa and M.~Savoji,
\newblock ``{MMSE speech enhancement based on GMM and solving an
  over-determined system of equations},''
\newblock in {\em Proc. IEEE Int. Symp. on Intell. Signal Process.}, Sep. 2011,
  pp. 1--5.

\bibitem{savoji2014speech}
M.~Savoji and S.~Chehrehsa,
\newblock ``{Speech enhancement using Gaussian mixture models, explicit
  Bayesian estimation and Wiener filtering},''
\newblock {\em Iranian J. of Electr. Electron. Eng.}, vol. 10, no. 3, pp.
  168--175, Sep. 2014.

\bibitem{ephraim1984speech}
Y.~Ephraim and D.~Malah,
\newblock ``Speech enhancement using a minimum-mean square error short-time
  spectral amplitude estimator,''
\newblock {\em IEEE Trans. on Audio, speech, and Lang. Process.}, vol. 32, no.
  6, pp. 1109--1121, Dec. 1984.

\bibitem{mohammadiha2013supervised}
N.~Mohammadiha, P.~Smaragdis, and A.~Leijon,
\newblock ``Supervised and unsupervised speech enhancement using nonnegative
  matrix factorization,''
\newblock {\em IEEE Trans. on Audio, speech, and Lang. Process.}, vol. 21, no.
  10, pp. 2140--2151, Jun. 2013.

\bibitem{wang2014training}
Y.~Wang, A.~Narayanan, and D.~Wang,
\newblock ``On training targets for supervised speech separation,''
\newblock {\em IEEE Trans. on Audio, speech, and Lang. Process.}, vol. 22, no.
  12, pp. 1849--1858, Aug. 2014.

\bibitem{wang2005ideal}
D.~Wang,
\newblock ``On ideal binary mask as the computational goal of auditory scene
  analysis,''
\newblock in {\em Speech separation by humans and machines}, pp. 181--197.
  Springer, 2005.

\bibitem{williamson2015complex}
D.~S. Williamson, Y.~Wang, and D.~Wang,
\newblock ``Complex ratio masking for monaural speech separation,''
\newblock {\em IEEE Trans. on Audio, speech, and Lang. Process.}, vol. 24, no.
  3, pp. 483--492, Dec. 2015.

\bibitem{choi2018phase}
H.~S. Choi, J.~H. Kim, J.~Huh, A.~Kim, J.~W. Ha, and K.~Lee,
\newblock ``Phase-aware speech enhancement with deep complex u-net,''
\newblock in {\em Proc. IEEE Int. Conf. on Learn. Representations}, Sep. 2018.

\bibitem{pascual2017segan}
S.~Pascual, A.~Bonafonte, and J.~Serra,
\newblock ``{SEGAN: Speech enhancement generative adversarial network},''
\newblock in {\em Proc. INTERSPEECH}, 2017, pp. 3642--3646.

\bibitem{rethage2018wavenet}
D.~Rethage, J.~Pons, and X.~Serra,
\newblock ``A wavenet for speech denoising,''
\newblock in {\em Proc. IEEE Int. Conf. on Acoust., Speech and Signal
  Process.}, Apr. 2018, pp. 5069--5073.

\bibitem{luo2019conv}
Y.~Luo and N.~Mesgarani,
\newblock ``Conv-tasnet: Surpassing ideal time--frequency magnitude masking for
  speech separation,''
\newblock {\em IEEE Trans. on Audio, speech, and Lang. Process.}, vol. 27, no.
  8, pp. 1256--1266, May 2019.

\bibitem{pandey2019tcnn}
A.~Pandey and D.~Wang,
\newblock ``{TCNN: Temporal convolutional neural network for real-time speech
  enhancement in the time domain},''
\newblock in {\em Proc. IEEE Int. Conf. on Acoust., Speech and Signal
  Process.}, May 2019, pp. 6875--6879.

\bibitem{zeghidour2021wavesplit}
N.~Zeghidour and D.~Grangier,
\newblock ``{Wavesplit: End-to-end speech separation by speaker clustering},''
\newblock {\em IEEE Trans. on Audio, speech, and Lang. Process.}, vol. 29, pp.
  2840--2849, Jul. 2021.

\bibitem{saleem2019deep}
N.~Saleem, M.~Irfan Khattak, M.~Y. Ali, and M.~Shafi,
\newblock ``Deep neural network for supervised single-channel speech
  enhancement,''
\newblock {\em Arch. Acoust.}, vol. 44, 2019.

\bibitem{haykin2008adaptive}
S.~S. Haykin,
\newblock {\em Adaptive filter theory},
\newblock Pearson Education India, 2008.

\bibitem{lim1979enhancement}
J.~S. Lim and A.~V. Oppenheim,
\newblock ``Enhancement and bandwidth compression of noisy speech,''
\newblock vol. 67, no. 12, pp. 1586--1604, 1979.

\bibitem{vaseghi2008advanced}
S.~V. Vaseghi,
\newblock {\em Advanced digital signal processing and noise reduction},
\newblock John Wiley \& Sons, 2008.

\bibitem{garofolo1993timit}
J.~S. Garofolo,
\newblock ``Timit acoustic phonetic continuous speech corpus,''
\newblock {\em Linguistic Data Consortium, 1993}, 1993.

\bibitem{Noisex}
{The Rice University, Signal Processing Information Base (SPIB)},
\newblock ``{Noise Data},'' 1995.

\bibitem{recommendation2001perceptual}
ITU-T,
\newblock ``Perceptual evaluation of speech quality (pesq): An objective method
  for end-to-end speech quality assessment of narrow-band telephone networks
  and speech codecs,''
\newblock {\em International Telecommunication Union, Recommendation P.862},
  2001.

\bibitem{taal2011algorithm}
C.~H. Taal, R.~C. Hendriks, R.~Heusdens, and J.~Jensen,
\newblock ``An algorithm for intelligibility prediction of time--frequency
  weighted noisy speech,''
\newblock {\em IEEE Trans. on Audio, speech, and Lang. Process.}, vol. 19, no.
  7, pp. 2125--2136, Feb. 2011.

\end{thebibliography}

\end{document}